# Mind the Gap: Trade-Offs Between Distributed Ledger Technology Characteristics
# (Working Paper)

Niclas Kannengießer

    Institute of Applied Informatics and Formal Description Methods, Karlsruhe Institute of Technology, Karlsruhe, Germany, niclas.kannengiesser@kit.edu

    Blockchain Center EU, University of Kassel, Kassel, Germany

Sebastian Lins

    Institute of Applied Informatics and Formal Description Methods, Karlsruhe Institute of Technology, Karlsruhe, Germany, sebastian.lins@kit.edu

Tobias Dehling

    Institute of Applied Informatics and Formal Description Methods, Karlsruhe Institute of Technology, Karlsruhe, Germany, dehling@kit.edu

Ali Sunyaev

    Institute of Applied Informatics and Formal Description Methods, Karlsruhe Institute of Technology, Karlsruhe, Germany, sunyaev@kit.edu

## ABSTRACT

When developing peer-to-peer applications on Distributed Ledger Technology (DLT), a crucial decision is the selection of a suitable DLT design (e.g., Ethereum) because it is hard to change the underlying DLT design post hoc. To facilitate the selection of suitable DLT designs, we review DLT characteristics and identify trade-offs between them. Furthermore, we assess how DLT designs account for these trade-offs and we develop archetypes for DLT designs that cater to specific quality requirements. The main purpose of our article is to introduce scientific and practical audiences to the intricacies of DLT designs and to support development of viable applications on DLT.

## CCS CONCEPTS

•Networks~Network protocols~Application layer protocols~Peer-to-peer protocols

•Software and its engineering~Software creation and management~Designing software~Software design tradeoffs

# 1   Introduction

Distributed ledger technology (DLT) enables the operation of a highly-available, append-only database (a distributed ledger), which is maintained by physically-distributed storage and computing devices (referred to as nodes), in an untrustworthy environment. DLT promises to increase efficiency and transparency of collaborations between individuals and/or organizations based on inherent qualities such as tamper- and censorship resistance, and democratization of data [1]. As a consequence, an ever-increasing number of applications on DLT have been developed in various domains, such as supply chain management [2], finance [3], or health care [4]. In supply chain management, product provenance systems employ DLT, for example, as a tamper-resistant data storage that is replicated across multiple nodes of collaborating entities in the supply chain [5,6]. Applications use distributed ledgers as a shared infrastructure that facilitates, for instance, reliable and tamper-resistant data storage, processing of transactions (e.g., for the transfer of digital assets), and automation of business processes [7,8]. Each application on DLT builds upon a particular DLT design (e.g., Ethereum or IOTA) that is defined as an instantiation of a DLT concept (e.g., blockchain) with unique characteristics [9].

Despite the promising benefits of DLT, past implementations of applications on DLT reveal critical dependencies between DLT characteristics that result in trade-offs; that is, the improvement of one DLT characteristic interferes with another DLT characteristic. For example, a trade-off exists between achieving availability and consistency in distributed ledgers [10]. High availability of a distributed ledger can be achieved by increasing the number of replications of the ledger. As a consequence, the network of nodes in the distributed ledger increases, however, this leads to reduced consistency due to message propagation delays [11]. Given the prevalent trade-offs between DLT characteristics, there will be no one-size-fits-all DLT design for applications on DLT. Rather, there will be DLT designs that are specialized to fulfill certain requirements but perform poorly on other requirements (e.g., low throughput, poor scalability, or high cost) due to drawbacks resulting from DLT-inherent trade-offs [9,12]. It is thus challenging to select suitable DLT designs for a given application and to assess potential drawbacks for the respective application on DLT. Making careful and well-founded decisions in favor for a (suitable) DLT design to develop viable applications on DLT is even more crucial because technical differences between DLT designs (e.g., different data structures and consensus mechanisms) impede the migration of data between distributed ledgers [13]. In this context, viability refers to applications' ability to operate over a long period under consideration of potentially changing requirements or improvements and resulting updates. To understand the trade-offs between DLT characteristics and their impact on the viability of applications on DLT, a comprehensive analysis of dependencies between DLT characteristics and resulting trade-offs is required.

While the body of research on DLT was ever increasing in the last decade, related research on DLT characteristics predominantly focuses on assessing the importance of characteristics for particular use cases (e.g., cryptocurrencies [14]) and on comparing application requirements with capabilities of selected distributed ledgers [e.g., 15]. Prior analyses of dependencies between DLT characteristics only consider a sparse set of DLT characteristics (e.g., integrity or scalability [16,17]). In addition, research on DLT characteristics and their dependencies is largely scattered across disciplines and needs to be synthesized in order to obtain a comprehensive understanding of dependencies between DLT characteristics and resulting trade-offs that limit the applicability of DLT designs to certain applications on DLT. We therefore strive to answer the following research question:

*How do trade-offs between DLT characteristics impact the viability of applications on DLT?*

To answer our research question, we applied a three-step research approach. First, we identified prevalent DLT characteristics by conducting a comprehensive literature review comprising 191 articles, and surveying DLT experts. Second, we analyzed the identified DLT characteristics in detail to uncover trade-offs in DLT designs, which were then applied to the most fitting DLT designs. Finally, we consolidated the identified trade-offs into archetypes and derived implications for applications on DLT.

Our study identified a consolidated list of forty DLT characteristics that are fundamental for assessing the suitability of DLT designs for applications on DLT, which we grouped into six DLT properties. This manuscript uncovers and explains twenty-four trade-offs between DLT characteristics and discusses the resulting drawbacks for applications. The identified DLT characteristics and properties range from purely technical (e.g., *strength of cryptography* in *security*) to social (e.g., *degree of decentralization* in *policy*), which highlights the complexity of DLT. Finally, we consolidated our findings into six DLT archetypes that indicate benefits and drawbacks for applications on DLT resulting from choice and configuration of a DLT design optimized toward a certain DLT property.

Our work contributes to the development of viable applications on DLT by discussing benefits and drawbacks of applications on DLT and presenting six archetypes of DLT designs. This work forms a bridge between currently separated research streams on DLT and forms a foundation for research on suitability assessments of DLT designs for applications. Our work allows practitioners and researchers to better understand which drawbacks for applications on DLT result from what configurations of DLT characteristics. Overall, we contribute to the scientific knowledge base by making it possible to set DLT characteristics into relation with applications on DLT and vice versa.

The manuscript is structured as follows. First, we introduce the current state of research on DLT and outline smart contract vulnerabilities and several attacks on distributed ledgers. This knowledge is required to understand the origins of trade-offs between DLT characteristics and drawbacks for applications on DLT. Second, we describe the method applied. Third, we present the identified DLT characteristics, the derived trade-offs between DLT characteristics, and the generated archetypes. Finally, we discuss our principle findings, summarize the implications for both practice and research, discuss research limitations, and give an outlook for future research.

## 2 Research Background and Related Research

### 2.1 Distributed Ledger Technology

In its essence, DLT serves as a shared, digital infrastructure for applications on DLT (e.g., in financial transactions [18]) by enabling the operation of a highly available, append-only distributed database (referred to as distributed ledger) in an untrustworthy environment [19], where separated storage and computing devices (referred to as nodes) maintain a local replication of the ledger. Nodes are maintained and controlled by individuals or organizations (referred to as node controllers [1]). An untrustworthy environment is characterized by the arbitrary occurrence of Byzantine failures [20,21], such as crashed or (temporarily) unreachable nodes, network delays, and malicious behavior of nodes.

In DLT, data is transferred and appended to the ledger in the form of transactions and is stored in a chronologically-ordered sequence. Each transaction contains meta-data (e.g., transaction recipient or timestamp) and a digital representation of certain assets (e.g., coins) or program code of a smart contract (see Section 2.3) [22]. When a node receives a new transaction, the transaction is validated by a proof of ownership for the digital representation of the asset based on digital signatures and public key cryptography [22,23].

DLT covers various DLT concepts, DLT designs, DLT properties, and DLT characteristics [9,24] (see Figure 1). **DLT concepts** describe the basic structure and functioning of DLT designs on a high level of abstraction. For instance, blockchain is a DLT concept describing the use of blocks that form a linked list. Each block contains multiple transactions that have been added into the block by nodes. Blockchains mostly follow the concept of replicated state machines, where each node maintains a local replication of the ledger in a certain state $s_n$ with an incrementing counter $n \in \mathbb{N}_0$, which expresses the height of a ledger (also called *block height* in blockchain). Appending new blocks or transactions to the local replication of the ledger represents a transition from a state $s_n$ to the subsequent state $s_{n+1}$. For example, Alice owns 10 coins in $s_n$ and she sends 2 coins to Bob. The transaction is first put in a queue of transactions to be processed and is eventually committed to the ledger. This commit initiates a transition from $s_n$ to $s_{n+1}$, in which Alice has a new balance of 8 coins and Bob's balance is increased by 2 coins. Other DLT concepts do not rely on generating a single

---

[1] We prefer the term *node controller* to *node provider*, because the *node provider* could be a cloud service provider (e.g. in Blockchain as a Service) that only hosts the node, while the *node controller* could maliciously influence the behavior of the node.

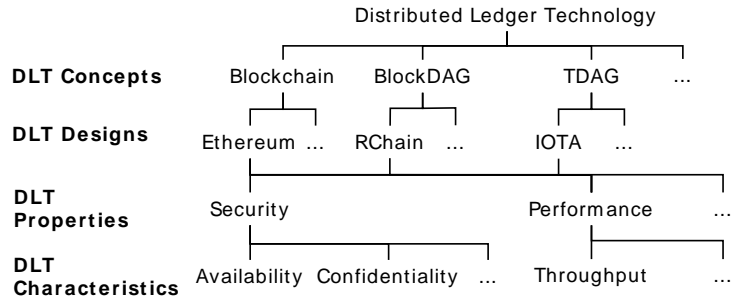

**Figure 1: Hierarchical structure of Distributed Ledger Technology (DLT), subordinate DLT concepts, and DLT designs and their respective DLT properties and DLT characteristics**

chain of blocks or even do not use blocks at all. Instead, for example, the DLT concept BlockDAG links generated blocks in a directed acyclic graph (DAG), while in a transaction-based DAGs (TDAG) transactions are linked directly with each other[2].

*DLT designs* specify the abstract description of DLT concepts by adding concrete values and processes for inherent DLT characteristics such as a maximum block size or a consensus mechanism to achieve a certain fault-tolerance. There are important differences between DLT designs that make DLT designs suitable for some applications and unsuitable for others. For instance, the DLT design Bitcoin creates a new block every 10 minutes and has a fixed, maximum block size of 1 MB [23]. In contrast, the DLT design Ethereum publishes new blocks every 17 seconds on average and block size is decided by individual nodes to increase flexibility of the distributed ledger. An instantiation of the formal specification of a DLT design is a distributed ledger.

*DLT characteristics* represent features of DLT designs, which are of technical (e.g., block creation interval) or administrative (e.g., node controller verification) nature, where the technical characteristics constrain future changes of the administrative characteristics (e.g., lack of scalability regarding network size of a distributed ledger). *DLT properties* are groups of DLT characteristics and shared by each DLT design. For instance, throughput and scalability are both associated with the DLT property performance. Although all DLT designs cover all DLT properties, DLT designs must not cover all DLT characteristics. For instance, TDAGs do not use blocks and do not feature any DLT characteristics related to blocks (e.g., block size, block creation interval).

All nodes of a distributed ledger maintain a local replication of the ledger, which is why all nodes must be synchronized and agree on a common state of the distributed ledger to reach consistency (e.g., agreeing that Bob's balance increased after receiving coins from Alice). For this purpose a consensus mechanism is employed to manage the negotiation between nodes, which (eventually) agree on a common state of the ledger [23,27]. Consensus mechanisms build upon trust models, which consider threats and uncertainties in the process of consensus finding such as Byzantine failures. Trust models form a set of assumptions, which must hold to assure consensus finding among nodes (e.g., at least 51 % of nodes must agree on a certain state). In Bitcoin, the first Byzantine fault-tolerant consensus mechanism that can be applied on a large scale and is able to prevent double spending (see Section 2.2) was presented: the Proof of Work (PoW)-based Nakamoto consensus [23]. Nevertheless, Nakamoto consensus comes with several drawbacks, such as poor throughput, exhaustive energy consumption, and vulnerability to attacks on integrity (see Section 2.2). To overcome drawbacks of the Nakamoto consensus, numerous alternative consensus mechanisms have been developed and already applied to DLT designs, such as GoChain [28], Hyperledger Fabric [29], soteriaDAG [26], and Tendermint [30]. In addition, BlockDAGs and TDAGs often employ alternatives to replicated state machines in their consensus mechanism, where not all nodes need to maintain an identical replication of the ledger. Such alternatives make use of random walks (e.g., in IOTA), clustering (e.g., in seele), or only keep transactions of a certain user on individual nodes (e.g., Nano). The consensus mechanisms discussed in this work are summarized in Table 1.

---

[2] Although blockchain represents a special type of BlockDAG, we decided to separate blockchain from BlockDAGs because of the different validation processes, data structures, and block storage organization [22,23,25,26]. While in blockchain all nodes work on the same block and only one block is appended to the blockchain, in BlockDAGs nodes work on different blocks that are added in parallel.

**Table 1: Selection of relevant consensus mechanisms for this work**

| Consensus Mechanism | Identifier | DLT Concept | Finality | Exemplary DLT Designs |
|---|---|---|---|---|
| CBC Casper | Casper | BlockDAG | Probabilistic | RChain [25] |
| Delegated Proof of Stake | DPoS | Blockchain | Total | EOS [31] |
| Delegated Proof of Stake | DPoS | TDAG | Probabilistic | Nano [32] |
| Modified Nakamoto Consensus using Greedy Heaviest Observed Sub Tree (GHOST) | PoW | Blockchain | Probabilistic | Ethereum [22] |
| Nakamoto Consensus | PoW | Blockchain | Probabilistic | Bitcoin [23] |
| PHANTOM | PHANTOM | BlockDAG | Probabilistic | soteriaDAG [26] |
| Practical Byzantine Fault Tolerance | PBFT | Blockchain | Total | Hyperledger Fabric [33] |
| Proof of Authority | PoA | Blockchain | Total | Ethereum [34] |
| Proof of Elapsed Time | PoET | Blockchain | Probabilistic | Hyperledger Sawtooth [35] |
| Proof of Reputation | PoR | Blockchain | Total | GoChain [36] |
| Proof of Stake | PoS | Blockchain | Probabilistic | Dash [37] |
| Tendermint Core | Tendermint | Blockchain | Total | Tendermint [30] |
| Tangle | Tangle | TDAG | Probabilistic | IOTA [38] |

In large, distributed ledgers (e.g., Bitcoin or Ethereum), where nodes can arbitrarily join and leave the network, it is not possible to reach consensus among all nodes before new data is added to the ledger [39]. Thus, newly appended data is not finalized and only *probabilistic finality* is given; that is, the data cannot be altered or removed with a certain probability [40]. The probability of finality of a transaction increases with more blocks (or transactions) that are appended to the distributed ledger after the transaction. Accordingly, the trust model of probabilistically final DLT designs (e.g., Bitcoin) allows for network partitions by design. Some nodes may agree on a state $s_{n,1}$ and others agree on $s_{n,2}$ with $s_{n,1} \neq s_{n,2}$. Network partitions which maintain different states are called *forks*. There can be an arbitrary number of forks in a distributed ledger and the DLT design needs to apply a rule to decide on a block (or transaction) being included into the main branch of the ledger and the ones not being part of it (named *stale blocks* or *stale transactions*). Fork resolution rules determine a certain state of the ledger to be correct, thereby, returning the system to a consistent state. In contrast to probabilistic finality, there is *total finality* (or just finality), where all nodes agree on the new state before data is appended to the ledger [41]. Once appended, data cannot be altered or removed anymore and forks such as in Bitcoin or Ethereum are not even possible (see Table 1).

Despite the widespread distinction between public and permissioned DLT design [e.g., 42,43] or public, consortium, and private DLT design [e.g., 44], we use a more granular terminology to make the trade-offs in the following sections unambiguous (in line with [15]). We distinguish between public and private DLT designs depending on the fact if a new node can directly join a network (referred to as public DLT) or if a permission must be granted first (referred to as private DLT). The distinction into public-private refers to read permissions and can be further distinguished into permissionless and permissioned, which refers to write permissions. Nodes can either all have the same permission (referred to permissionless) or must first be granted permission to validate and commit new data (referred to permissioned). The used terminology is summarized in Table 2.

**Table 2: Exemplary classification of DLT designs according to the used terminology and the respective focus**

|  | **Public** | **Private** |
|---|---|---|
| **Permissioned** | GoChain [28] *High performance general purpose* | Quorum [45] *Financial asset transfers* |
| **Permissionless** | Ethereum [22] *General purpose* | ARK Ecosystem [46] *Flexibility for developers* |

In *public-permissionless* DLT designs (e.g., Bitcoin), an incentive mechanism is required because validating nodes must be motivated to share their computational resources. The incentive mechanism specifies a reward scheme for nodes that participate in the generation and/or validation of blocks and transactions, consensus finding, and maintenance of the distributed ledger. The participation of nodes in a distributed ledger to receive a monetary reward is called mining. Accordingly, validating nodes are often referred to as miners. For example, validating nodes in the Bitcoin network receive a certain amount of coins if they are the first to create a valid new block. Such incentive mechanisms are predominantly applied to distributed ledgers that employ nodes of unknown node controllers, thus, allow for a high degree of decentralization.

A distributed ledger's *degree of decentralization* refers to the number of independent validating node controllers reduced by the number of controllers that control more than average validating nodes divided by the total number of nodes in the DLT network. Consequently, a distributed ledger's degree of decentralization is determined by two dimensions: the number of independent validating node controllers (e.g., companies or individuals) and the number of validating nodes (see Figure 2). If the number of validating nodes increases, and all additional nodes are maintained by the same controller, the degree of decentralization would decrease because this controller gains unproportionally influence on the distributed ledger's consensus finding and integrity. On the contrary, the degree of decentralization is increased as independent node controllers add nodes of at most average computational resources to the distributed ledger.

## 2.2 Attack Vectors and Vulnerabilities

To understand drawbacks of applications on DLT in the form of vulnerabilities, it is important to introduce potential attack vectors. Although DLT is often considered to be immutable, there have already been millions of dollars lost due to successful attacks on distributed ledgers that rewrote the transaction history (e.g., 51% Attack [47]). In this section, we explain the most prominent attacks on the integrity of a DLT design, which play a role in the identified trade-offs. It should be noted that the explained attacks predominantly target forkable DLT designs (e.g., Bitcoin or Ethereum) because there is only little research on the security of DAGs.

*Double Spending.* Double spending refers to multiple use of a particular asset by the same user for different purposes without the asset being returned before using it again [23]. In a double spending attack, the attacker suggests to a user that a product was paid on a certain network partition visible for the user, while transferring the just issued coins back to her own wallet on another network partition. After the network partitions were resolved, the attacker still owns her coins and the product she actually did not pay for.

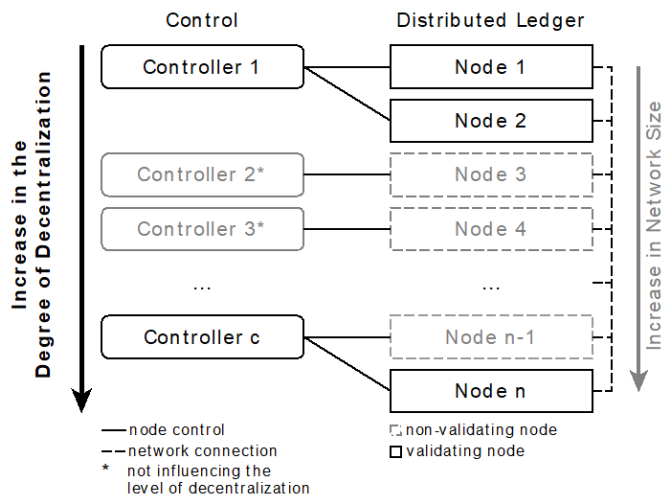

Figure 2: The degree of decentralization is determined by the number of independent validating node controllers (e.g., an organization or individuum) and the number of validating nodes in the distributed ledger. An increase in number of independent controllers who control validating nodes increases the overall degree of decentralization of a distributed ledger.

*Partition-based Attacks.* Partition-based attacks can be successfully performed in *public-permissionless* DLT designs with probabilistic finality (e.g., Bitcoin and Ethereum). In such DLT designs, forks can be exploited to perform attacks on the distributed ledger's integrity. The most popular attacks are *51 % attacks*, *balance attacks*, and *eclipse attacks*. A **51 % attack** can be successful in DLT designs with a consensus mechanism that relies on a majority decision among nodes (e.g., Nakamoto Consensus). If the attackers control the majority of nodes, they can rewrite the transaction history because their majority of nodes agrees on their desired (fraudulent) state of the distributed ledger. In DLT designs such as Bitcoin, where nodes can arbitrarily join and leave the network, mechanisms are required to prevent attackers from setting up a huge number of virtual nodes. Such mechanisms usually employ proof of work [48], where nodes must first do computational work before new data can be committed to the distributed ledger. A **balance attack** incorporates the process of transiently disrupting communications between subgroups of validating nodes with equal computational power, which is determined by, for example, the nodes' hashing rate [49]. While the communication is disrupted, transactions can be submitted to one subgroup while the attacker mines in another subgroup. The attackers' aim is to outweigh the blockchain branch they submitted transactions to with the blockchain branch they work on to rise probability for successful double spending. As a result, the ledger may be rewritten at any time the attacker prefers [49]. In an *eclipse attack*, attackers target network partitions by delaying or blocking message forwarding (e.g., transactions) to nodes of the attacked network partition [50]. Due to the delay of messages, targeted nodes are isolated from the network. Such network partitions facilitate double spending. For example, an attacking node would send a transaction (e.g., a payment) to the victim node. The victim node validates the transaction and is, subsequently, eclipsed from the network by the attacker. Then, the attacker issues another transaction to the entire network and spends the same assets again. Since all partition-based attacks target information asymmetry among nodes, partition-based attacks can be successfully performed in combination with *routing attacks* (i.e., Border Gateway Protocol hijack attack [51]). In routing attacks, attackers manipulate nodes or network operators to intentionally delay network messages [51].

*Bribery Attack.* During a bribery attack, an attacker strives to create a new main branch by incentivizing validating nodes to work on a particular fork of the DLT design, which the attacker rules [52]. Thus, the number of nodes that work on the attacker's branch increases and the attacker's branch may eventually catch up with the main branch of the blockchain and, finally, become the main branch.

*Selfish-Mining.* Selfish-mining attacks describe a phenomenon where a set of nodes work on their own branch of a blockchain without publishing their blocks to the main branch until their branch would be chosen as future main branch by the fork resolution rule applied to the distributed ledger [16,53]. A selfish-mining attack is carried out by attackers to obtain excessive rewards or waste the computing power of honest validating nodes [53]. It was found that a successful selfish-mining attack can be performed in Bitcoin if at least one third of the validating nodes collude [16].

*Long-Range Attack.* A long-range attack aims to rewrite the transaction history from the genesis block (the first block in a blockchain). A long-range attack is successful when the attacker has generated a fork, which has become the main chain, which is similar to selfish-mining (or short-range attack) [54]. Predominantly DLT designs, which rely on Proof of Stake as a consensus mechanism are prone to long-range attacks because PoW-based DLT designs require too much computational effort to rewrite transaction history beginning from the genesis block, which is why long-range attacks are considered impractical for PoW-based DLT designs [54].

*Blockchain Anomaly.* The blockchain anomaly refers to the fact that a blockchain cannot guarantee that a committed transaction is permanently included in a fork of a blockchain. Due to this, conditional transactions are hard to perform. In conditional transactions, a transaction $t_{i+1}$ of a node $n_1$ should be committed after a certain condition is fulfilled (e.g., the commit of a previous transaction $t_i$ issued by $n_2$) [27]. If the transactions have been issued by $n_1$ and $n_2$ to different network partitions, it is likely that $t_i$ and $t_{i+1}$ are included in different forks of the blockchain. The blockchain may finally decide for the fork not containing $t_i$ but $t_{i+1}$ and only commit $t_{i+1}$, which violates the conditional execution of $t_i$ and $t_{i+1}$. The blockchain anomaly can occur in blockchains whose consensus mechanism does not ensure deterministic agreement between nodes and consensus safety [27].

*Sybil Attack.* In a sibyl attack, the attacker sets up multiple (virtual) nodes to contribute the majority of actors in consensus finding to eventually rewrite the transaction history of the distributed ledger. To decrease the probability of successful sibyl attacks, all nodes must perform a certain PoW, where each node must first finish a computationally hard task, which can easily be evaluated by other nodes [48]. For example, in Bitcoin, Ethereum, and Nano, the block (or transaction) issuer must first guess a nonce with a corresponding hash value, which fulfills an easy to validate condition (e.g., starting with a defined minimum number of zeroes). Sybil attacks can be performed to isolate (honest) nodes from the network by not relaying transactions of these nodes [55,56]. The selective relaying of transactions can contribute to double spending [57].

## 2.3 Smart Contracts and Respective Vulnerabilities

Several distributed ledgers offer the possibility to deploy and execute customized business logic through smart contracts. Smart contracts are software programs, which are developed in basic OP_CODE (e.g., in Bitcoin Script language) [58,59] or in high-level programming languages (e.g., Java, Python, or Solidity), allowing for Turing completeness [22,31,60]. When Bitcoin was invented, the development of smart contracts was limited to the use of cryptographic functions such as hash-locks, time-locks, and multi-signatures. To give more flexibility to developers, the Ethereum foundation developed the Ethereum Virtual Machine (EVM) that allows for the execution of Turing complete smart contracts, which can be developed in high-level programming languages such as Solidity [22]. An Ethereum smart contract is contained in a transaction, which is sent to and eventually stored on the Ethereum blockchain. Ethereum smart contracts can receive and keep assets and issue transactions. The smart contract can be called via its unique address to trigger methods [22]. If an Ethereum smart contract is triggered by a transaction, each node of the distributed ledger separately executes the smart contract. Smart contract computations are not restricted to the use of data already stored on the ledger (on-chain data), but can also retrieve data from external data feeds (referred to as oracles) and outsource computation-heavy processes [61].

Smart contracts are of high interest in the field of DLT because they enable the development of applications on DLT. However, smart contracts leverage programming paradigms that developers are not yet used to (e.g., rollbacks of failed transactions due to out-of-gas conditions). Since Ethereum introduced Turing complete smart contracts on *public-permissionless* DLT designs, the issue of how to prevent infinite loops became crucial to prevent system failure. As a solution, a pricing schedule, which requires an economic equivalent (referred to as gas in Ethereum) to be paid for the execution of a particular smart contract, is applied in *public-permissionless* DLT designs [22]. As soon as the quantity of gas is no longer sufficient to execute a smart contract, its execution is cancelled (out-of-gas condition). Out-of-gas conditions always require appropriate error handling. Otherwise, the respective smart contracts is locked automatically and cannot be executed anymore [62,63]. In the following, we briefly review the smart contract vulnerabilities relevant for this work.

*Overflow/Underflow.* Numbers in smart contracts, especially, those being executed in the EVM, are usually stored in variables of the datatype unsigned integer (*uint*). If the stored values exceed the maximum *uint* value (overflow), the value is set to zero. If the value of a *uint* variable becomes smaller than zero (underflow) it is set to its maximum value [62,64]. Attacks can exploit over- and underflows for different purposes such as manipulation of payout values. To prevent overflow/underflow attacks, developers must consider whether the *uint* value could exceed its maximum or become less than zero.

*Unbounded Loops.* The most standard form of a gas-focused vulnerability is that of unbounded loops. Loops whose behavior is determined by user input could iterate too many times, exceeding the block gas limit, becoming too expensive economically to perform, or lead to overflow or underflow. For example, a list could become a cause for an unbounded loop if users can add arbitrary entries and, thus, increment the number of iterations necessary to go through the list, where each iteration costs gas. This will commonly lead to a *Denial of Service* for all transactions that must attempt to iterate the loop [62].

***Reentrancy.*** Atomicity and sequentiality of transaction execution require that non-recursive methods cannot be re-entered before their return values are committed to memory. The requirements for atomicity and sequentiality are by default not fulfilled in smart contracts and must be considered by smart contract developers. Recursive calls of smart contract functions (referred to as reentrancy) can occur as a single smart contract invokes itself or in a chained execution of smart contracts. Often, such recursive calls neglect the execution model underlying smart contracts (e.g., finite state machines), where each change in the smart contract's data represents a transition to a new state of the smart contract. The execution model allows for the execution of functions (e.g., withdraw functions) without changing the smart contracts internal state. [64,65]. In one of the most prominent incidents in the context of smart contracts, the attack on the Decentralized Autonomous Organization (The DAO) [65], reentrancy was exploited, which caused the hard fork of Ethereum into Ethereum Classic and Ethereum in 2016.

***Wallet Griefing.*** A smart contract can cause unexpected errors when invoking external methods that may itself throw an out-of-gas exception [66]. In the EVM transactions are, for example, issued to an account using the *<recipientAccount>.send(uint)* function. Using this function to transfer tokens can lock the smart contract if error handling is not properly implemented because the execution of *<recipientAccount>.send(uint)* can produce out-of-gas conditions [66]. Wallet griefing is also realistic when the smart contract should handle multiple clients without isolation and when a failure in sending transactions using *<recipientAccount>.send(uint)* occurs [62].

## 2.4 Prior Research on Trade-Offs between DLT Characteristics

Viability and maintainability of applications on DLT heavily depends on the choice of a DLT design. Maintainability refers to making an application easy to update and adapt to changing requirements. As DLT combines insights from several disciplines of computer science (e.g., distributed systems and cryptography) and economics (e.g., game theory), DLT designs are complex and implications for applications on the respective DLT design are not trivial to derive. Extant research on DLT can be distinguished into four research streams: *description, analysis, application,* and *guidance* (see Table 3).

The *description* research stream focuses on structured descriptions and classifications of DLT, for example, taxonomies of DLT characteristics. Characteristics of and differences between DLT designs are collected and consolidated into structured overviews [e.g., 14,67–69]. However, dependencies between identified DLT characteristics are seldom investigated and the causes for the ever-increasing number of DLT designs remain unclear. Hence, the practical or technical use of identified DLT characteristics for a comprehensive understanding of functionalities and constraints of DLT designs is limited to the provision of a common understanding of selected DLT designs, while implications for application development remain unclear.

In the second research stream, *analysis*, dependencies between selected DLT characteristics are measured and individual dependencies between DLT characteristics are reported. For example, high performance of a DLT design mostly comes at the cost of its level of security [11,53]. Extant research explains this trade-off in blockchains by the fact that various attacks result from an increased stale block rate, which is influenced, among other things, by the (mis-)configuration of block size and block creation interval in *public-permissionless* DLT designs [11,53]. However, the application perspective is not considered in prior analysis research because most of the research articles do not explain practical implications that result from the observed effects produced by configurations of DLT characteristics. Additionally, only few DLT characteristics have been included in these analyses (e.g., block size, throughput, or scalability). A holistic view on dependencies between DLT characteristics and resulting drawbacks is not presented in this stream.

**Table 3: Overview of prevalent DLT research streams**

| Stream | Description | Example |
|---|---|---|
| Description | Generation of structured descriptions and classifications of DLT designs | [14,69] |
| Analysis | Measurement and report of dependencies between particular DLT characteristics | [11,53] |
| Application | Development of prototypes and investigating the application of DLT designs in certain domains | [12,70] |
| Guidance | Development of processes to guide practitioners when looking for a suitable DLT design for applications | [69,71] |

The third research stream focuses on the *application* of DLT in certain domains, for example, supply chain management, health IT, or the Internet of Things. Due to the novelty of DLT, the potential for and usefulness of applications on DLT in different domains is still under investigation. Hence, several application prototypes on DLT were already developed and drawbacks due to the chosen DLT designs have been identified. For example, the Ethereum blockchain is considered to have a low throughput [12,70] and is costly [72] when used in the Internet of Things [70]. The Bitcoin blockchain cannot provide confidentiality and has an even lower throughput than Ethereum [73]. IOTA, which is predominantly designed for the use in the Internet of Things, is considered to be slow when handling a massive amount of data [74]. Hyperledger Fabric and Ripple [75] come with high throughput but limited scalability in the number of validating nodes [76]. The practical drawbacks caused by DLT designs are often mentioned in the application research stream, but the causes of the respective drawbacks are not further investigated.

The forth research stream, *guidance*, focuses on the development of processes to guide practitioners when looking for a suitable DLT design for applications. However, the presented processes are highly abstract and generic, and focus on questions related to whether a distributed ledger is useful at all. Some articles consider select DLT designs and compare them but hardly address causes for the viability of investigated DLT designs for applications [e.g., 17,77,78]. Other articles address the degree of decentralization [e.g., 69,71]. The technical fundamentals of DLT that are crucial for the viability of a DLT design for an application are only sparsely discussed in the guidance research steam. Therefore, existing measures to evaluate suitability of an underlying DLT design for an application cannot be effectively used and the assessment of drawbacks of applications on a particular distributed ledger remains unclear.

These four research streams provide valuable contributions in general and for identifying trade-offs in particular. While already some trade-offs have been identified in prior research, these research streams are disjunct, which is why it is hard to obtain a holistic overview of the implications of a DLT design for an application on DLT. More comprehensive analyses of trade-offs in DLT in extant research are limited to the context of electronic health records and consider only blockchains [9,79]. The findings in extant research on dependencies between DLT characteristics should be synthesized to identify trade-offs and support the development of viable applications on DLT for various use cases. This is the objective of our work.

## 3 Method

To answer our research question, *how do trade-offs between DLT characteristics impact the viability of applications on DLT*, we applied a three-step research approach. First, we identified prevalent DLT characteristics by conducting a descriptive literature review [80–82] and surveying DLT experts. Second, we analyzed the identified DLT characteristics in detail to uncover trade-offs in DLT designs. Finally, we consolidated the identified trade-offs into archetypes and derived implications for and drawbacks of applications on DLT.

### 3.1 Identification of DLT Characteristics

Our descriptive literature review [83] was guided by extant recommendations for literature reviews [84–86]. To identify publications addressing DLT characteristics, we searched scientific databases, which cover the top computer science conferences and journals: ACM Digital Library, EBSCOhost, IEEE Xplore, ProQuest, and ScienceDirect. To cover a broad set of publications, we searched each database with the following string in title, abstract, and keywords: *(blockchain\* OR ("distributed ledger\*"))*. We limited our search to peer-reviewed articles to ensure a high quality of articles. Our search in June 2018 identified 1,144 articles. To identify and filter articles, we first checked the relevance of each article by analyzing title, abstract, and keywords. If any indication for relevance appeared, the article was marked for further analysis. We excluded articles that were duplicates (62), grey literature (i.e., editorials, work-in-progress, dissertations) and books (18), not applicable to our study (56), or not available in English (31). This first relevancy assessment resulted in a sample of 977 potentially relevant articles. Afterwards, a fine-grained relevance validation was made by accessing and reading the article abstracts, resulting in a final sample of 191 relevant articles. In this second relevance assessment, we excluded non-research articles (76) and articles that did not relate to viability of DLT designs for applications on DLT (710).

After the literature search was completed, we, first, carefully read and analyzed the 191 articles to identify DLT characteristics. We recorded for each extracted DLT characteristic a name, a description, and the original source [87]. In total, 277 DLT characteristics were extracted. A list of master variables was created to aggregate the identified DLT characteristics. A master variable is an aggregation of similar DLT characteristics consisting of a master variable name and a master variable description [87]. If an identified DLT characteristic fitted into an existing master variable, we assigned it accordingly; otherwise, a new master variable was created. For example, we aggregated the DLT characteristics *immutability* and *tamper-resistance* to the master variable *integrity*. Since different people often put the same labels on different things, and vice versa, we considered semantic ambiguities (e.g., different terms for the same characteristic) during our data analysis [88]. To improve readability of this article, we use the term DLT characteristic for the identified master variables in the remainder of this manuscript because master variables represent aggregations of similar DLT characteristics. To ensure that we identified a reliable set of master variables, we aimed to reach theoretical saturation [89,90] with respect to the emerging DLT characteristics. Since no new master variable emerged in the last 27 articles identified in our literature review, we are confident to have reached theoretical saturation.

DLT characteristics are further grouped into DLT properties under consideration of their influence on the DLT design (e.g., performance or security). For instance, DLT characteristics were grouped into the DLT property *security* if they were related to common security topics such as *confidentiality, integrity,* and *availability*.

To consolidate and critically evaluate the derived DLT characteristics and DLT properties and their respective definitions, we set up an online survey to obtain feedback. We sent 68 requests for feedback via email to DLT experts who had at least three years of experience in dealing with DLT in a business or private context. 35 DLT experts participated in the survey and we received 113 comments on the generated DLT properties and DLT characteristics. We revised the DLT characteristics and DLT properties and their definitions according to the gathered feedback. For example, the DLT characteristic *cost* was split into *resource consumption* and *transaction fee*.

## 3.2 Uncovering Trade-Offs Between DLT Characteristics

We extracted dependencies between DLT characteristics described in the examined research articles. In case these research articles discuss relevant trade-offs, we coded trade-offs between DLT characteristics [e.g., 9,11,79]. In addition, we analyzed the identified dependencies between DLT characteristics (e.g., more replications of the stored data increase availability) and abstracted trade-offs (e.g., more replications increase the latency until consistency among all nodes is reached).

We evaluated the derived trade-offs on seven DLT designs including Bitcoin, Ethereum, and Hyperledger Fabric representing the blockchain concept; RChain and soteriaDAG representing the BlockDAG concept; and IOTA and Nano representing the TDAG concept (see Figure 1). In particular, we discussed the occurrence of the identified trade-offs in thoughtful group discussions by two authors and two PhD students having profound knowledge and experience in the domain of DTL. Prior to the group discussions, the four participants individually rated each trade-off for the selected ledgers based on their knowledge and experience. In addition, each participant studied available documentation and white papers for the selected DLT designs. Individual rating results were consolidated and then actively discussed by participants until all conflicts were resolved and consensus (total finality) was reached. Table 6 summarizes our findings.

## Table 4: Identified DLT properties

| DLT Property | Description |
|---|---|
| Flexibility | The degrees of freedom in deploying applications on and customizing a DLT design |
| Opaqueness | The degree to which the use and operation of a DLT design cannot be tracked |
| Performance | The accomplishment of a given task on a distributed ledger under efficient use of computing resources and time |
| Policy | The ability to guide and verify the correct operation of a DLT design |
| Practicality | The extent to which users of a distributed ledger can achieve their goals with respect to social and socio-technical constraints of everyday practice |
| Security | The likelihood that functioning of the distributed ledger and stored data will not be compromised |

## 3.3 Configuration of DLT Archetypes

To make the derived trade-offs between DLT characteristics more tangible and evaluate their impact on applications on DLT, we jointly configured DLT archetypes for each DLT property. These archetypes describe how to configure DLT characteristics to achieve a certain DLT property (e.g., flexibility) while considering underlying trade-offs. To identify the archetypes, we reviewed identified DLT characteristics for each DLT property and selected trade-offs corresponding to DLT characteristics first. We then decided what DLT characteristic is preferred over another to achieve the DLT property with respect to each trade-off. For example, an adequate block size outweighs transaction fees (see trade-off *G.1*) to achieve the DLT property flexibility. Finally, for each archetype and its corresponding property we highlight drawbacks for applications on DLT that are caused by the underlying DLT design. We assumed that all characteristics that are more positively associated with the property assigned to the archetype should have a high value and accounted for the respective trade-offs. For the performance archetype, we assumed, for example, high scalability and throughput and analyzed the effects of this configuration on DLT characteristics of other DLT properties (e.g., availability within security).

## 4 Trade-Offs between DLT Characteristics

## 4.1 DLT Characteristics and Trade-Offs between DLT Characteristics

The literature review revealed 40 DLT characteristics that are relevant for the assessment of a DLT design's viability for an application on DLT. The 40 DLT characteristics are briefly presented and defined in Table 5. The grouping of the 40 DLT characteristics resulted in a final set of 6 DLT properties, which are presented in Table 4. In the following, we will discuss derived trade-offs between DLT characteristics. Table 6 lists identified trade-offs.

**A Flexibility vs. Performance**

*A.1 Turing-complete Smart Contracts vs. Resource Consumption*
The use of external services in smart contracts via oracles enables more flexibility in defining the conditions that must be fulfilled before the smart contract issues transactions. If an oracle is requested from a smart contract, the oracle receives requests from every node because every node needs to execute the smart contract. Thus, an oracle can become a performance bottleneck because the oracles' bandwidth may not be sufficient to handle the amount of (almost) simultaneous requests by nodes.

*A.2 Turing-complete Smart Contracts vs. Transaction Fee*

In Bitcoin, no Turing-complete smart contracts can be developed and time complexity for processing a transaction (e.g., for multi-signature transactions) equals $k * N$ at maximum, where a transaction's payload incorporates $N$ bytes and a constant factor $k$. Due to the limited flexibility in Bitcoin smart contracts, there is no need to apply a mechanism to interrupt potential infinite loops (e.g., like *gas* in Ethereum). Instead, in Bitcoin transaction fees are employed to incentivize validating nodes to prefer the validation of a transaction

## Table 5: Identified DLT characteristics

| DLT Property | DLT Characteristic | Description |
|---|---|---|
| Flexibility | Interoperability | The ability to interact between DLT designs and with other external data services |
| | Maintainability | The degree of effectiveness and efficiency with which a DLT design can be kept operational |
| | Turing-complete Smart Contracts | The support of Turing-complete smart contracts within a DLT design |
| | Token Support | The possible uses of tokens within a DLT design (e.g., security token, stable coin, or utility token) |
| | Transaction Payload | The size of the payload in a transaction |
| Opaqueness | Traceability | The extent to which transaction payloads (e.g., assets) can be traced chronologically in a DLT design |
| | Transaction Content Visibility | The ability to view the content of a transaction in a DLT design |
| | User Unidentifiability | The difficulty of mapping senders and recipients in transactions to identities |
| | Node Controller Verification | The extent to which the identity of validating node controllers is verified prior to joining a distributed ledger |
| Policy | Auditability | The degree to which an independent third party (e.g., state institution, certification authority) can assess the functionality of a DLT design |
| | Compliance | The alignment of a DLT design and its operation with policy requirements (e.g., regulations or industry standards) |
| | Degree of Decentralization | The plurality of independent validating node controllers reduced by the number of controllers that control more than average validating nodes |
| | Incentive Mechanism | A structure in place to motivate node behavior that ensures viable long-term operation of a distributed ledger (e.g., by contributing computational resources) |
| | Liability | The existence of a natural or legal person that can be subjected to litigation with respect to the DLT design |
| Performance | Block Creation Interval | The time between the creation of consecutive blocks (only in DLT designs using blocks) |
| | Block Size Limit | The value of a fixed maximum storage size of a block (only in DLT designs using blocks) |
| | Confirmation Latency | The time span between the inclusion of a transaction in a ledger and the point in time where enough subsequent transactions have been included in the ledger so that the likelihood of future manipulations of the initial transaction becomes negligible |
| | Resource Consumption | The computational efforts required to operate a DLT design (e.g., for transaction validation, block creation, or storing the distributed ledger) |
| | Propagation Delay | The time between the submission of a transaction (or block) and its propagation to all nodes |
| | Scalability | The capability of a DLT design to efficiently handle decreasing or increasing amounts of required resources (e.g., of transactions per second or number of validating nodes) |
| | Stale Block Rate | The number of blocks that have been generated in a period of time but not appended to the main chain of the distributed ledger (only in forkable DLT designs using blocks) |
| | Throughput | The maximum number of transactions that can be appended to a DLT design in a given time interval |
| | Transaction Validation Latency | The time required for validating a transaction by validating nodes |
| Practicality | Transaction Fee | The price transaction initiators can or must pay for the processing of transactions |
| | Ease of Node Setup | The ease of configuring and adding a new or crashed node to the DLT design |
| | Ease of Use | The simplicity of accessing and working with a distributed ledger |
| | Support for Constrained Devices | The extent to which devices with limited computing capacities (e.g., sensor beacons) can participate in a DLT design |

## Table 5 cont.: Identified DLT characteristics

| DLT Property | DLT Characteristic | Description |
|---|---|---|
| Security | Atomicity | The state where transactions are either completely executed or not executed |
| Security | Authenticity | The degree to which the correctness of data that is stored on a distributed ledger can be verified |
| Security | Availability | The probability that a distributed ledger is operating correctly at any point in time |
| Security | Censorship Resistance | The probability that a transaction in a DLT design will be intentionally aborted by a third party or processed with malicious modifications |
| Security | Confidentiality | The degree to which unauthorized access to data is prevented |
| Security | Consistency | The absence of contradictions across the states of the ledger stored by all nodes participating in a DLT design |
| Security | Durability | The property that data committed to the ledger will not be lost |
| Security | Fault Tolerance | The constant maximum proportion of failed, malicious, or unpredictable nodes a DLT design can compensate while operating correctly |
| Security | Integrity | The degree to which transactions in the distributed ledger are protected against unauthorized (or unintended) modification or deletion |
| Security | Isolation | The property that transactions do not impact each other during their execution |
| Security | Non-Repudiation | The difficulty of denying participation in transactions |
| Security | Reliability | The ability of a system or component to perform its required functions under stated conditions for a specified time |
| Security | Strength of Cryptography | The difficulty of breaking the cryptographic algorithms used in the DLT design |

with higher transaction fees over transactions where the sender is only willing to pay a smaller transaction fee (e.g., in Bitcoin).

Turing completeness (e.g., in Ethereum) adds more flexibility to smart contracts but also increases complexity and vulnerabilities. Turing completeness allows for the use of loops in smart contract code, which may even result in infinite loops and, eventually, Distributed Denial-of-Service (DDoS) attacks. The (automated) detection of infinite loops is not possible due to the halting problem [91]. To cope with potential infinite loops in permissioned DLT designs, timeouts are often applied (e.g., in Hyperledger Fabric). Such timeouts, however, limit flexibility in the smart contract development because an upper boundary of time is defined to kill the execution of a smart contract. To overcome this limitation, a pricing schedule is applied in various DLT designs (e.g., *gas* in Ethereum; see Section 2.3) to incentivize validating nodes to execute smart contracts. Users must pay a certain charge for smart contract execution proportional to the smart contract's computational operations. In such distributed ledgers, transaction fees are thus both an incentive mechanism for nodes to process smart contracts and a security mechanism to prevent potential vulnerabilities stemming from Turing completeness in smart contracts.

*A.3 Turing-complete Smart Contracts vs. Transaction Validation Speed*

Support for more expressive programming languages (i.e., C++, Java, or Solidity) enables the development of smart contracts that offer a broad range of functionality. The more functionality is added to a smart contract, the higher becomes its complexity. Ultimately, this impedes performance because of the increased execution time for complex smart contracts. Consequently, the time required for transaction processing and validation increases [92].

**B Flexibility vs. Security**

*B.1 Maintainability vs. Availability*

To secure DLT designs, the software client of individual nodes must be maintainable and remain compatible with the majority of nodes in the network. Updates for the client protocol of a DLT design must be performed on each node. This is why maintainability of DLT designs decreases with an increasing number of independent nodes due to additional efforts when negotiating and applying software client updates. For example, in Bitcoin and LiteCoin, it has taken weeks to agree on updates such as the adoption of Segregated Witness (SegWit) and SegWit2x [93]. However, an increasing number of nodes increases the ledger's redundancy due to increasing replications, which is beyond what is possible with distributed databases prior to DLT. The dependency between maintenance-related cost (e.g., time and money) and the degree of decentralization of the distributed ledger is also depreciatingly known as blockchain bloat or DLT bloat [94].

*B.2 Maintainability vs. Integrity*

To allow for efficient maintenance of a distributed ledger, the coordination of update procedures should be facilitated by a low number of (independently controlled) nodes (see trade-off *B.1*). However, a decrease in the number of independently maintained nodes (hence, a decrease in the DLT design's degree of decentralization) impedes the integrity of DLT designs due to reduced absolute fault tolerance regarding the number of tolerable, malicious nodes.

On the other side, a high level of a distributed ledger's integrity also impacts maintainability of applications on DLT [95]. For achieving a high integrity of distributed ledgers, smart contracts are tamper-resistance: smart contracts must always be redeployed and initialized with the state of the obsolete version whenever the smart contract is updated. In addition, the stored smart contract addresses must be adapted in any module of the application and chained smart contract that references the deprecated smart contract. Hence, tamper-resistance and resulting integrity of a distributed ledger increases efforts for maintenance. However, by relaxing integrity and, thus, tamper-resistance of smart contracts, the idea of an inevitable and automated enforcement of agreements becomes vulnerable to malicious behavior.

*B.3 Turing-complete Smart Contracts vs. Confidentiality*

Use of smart contracts threatens confidentiality in three ways. First, it is publicly visible which account's transactions triggered a smart contract [96]. Second, the compiled smart contract code is also visible to the public and smart contracts can be decompiled to human readable source code. Thus, the current state of the smart contract and even values of variables that are declared private in the smart contract can be inferred due to the transparent smart contract code and transactions [96]. Hence, the common ways of using smart contracts do not support confidentiality. Nevertheless, there are new approaches for private smart contracts that tackle this issue. For example, as proposed in the HAWK framework [96], smart contracts can be divided into a private and a public part. The private part determines the payout distribution among involved parties; the input data (e.g., a number of coins) is kept private and is protected using zero-knowledge proofs [96]. As a result, no participant knows the input data other participants sent to the smart contract.[3] Third, oracles and/or external services might have insight into data that is exchanged via smart contracts. Such oracles and/or services are often centralized instances that forward certain data, for example, in Provable (formerly known as Oraclize) [98] or TownCrier [86]. The use of external services in DLT requires at least one trusted party, which stores the requested data. Thus, the oracle provider can have insights into data flows that are made by users who trigger a smart contract. Although requests may be (partially) executed in a protected SGX enclave[4] (e.g., Town Crier [86]), there is at least the risk of a leaked key that can be used to decrypt the respective data and the risk of failures in the centralized architecture.

**C Opaqueness vs. Performance**

*C.1 User Unidentifiability vs. Resource Consumption*

To achieve unidentifiability among users of a DLT design, additional computational resources are required such as computational power, storage space, and runtime [99]. For example, additional data structures can be used to increase unidentifiability, but require additional storage size [73,99,100] or zero knowledge proofs, which, for example, require various message exchanges between two parties to validate instead of one [101]. Thus, unidentifiability comes with the cost of high resource consumption.

---

[3] The public portion of a HAWK smart contract is composed of three parts: publicly executed code, privately executed code, HAWK manager code. While the publicly executed code is executed on each node of the distributed ledger, the privately executed portion of the smart contract code is only executed by users, who sent transactions to the smart contract. The HAWK manager is a trusted party, who runs the HAWK manager code in an Intel SGX enclave [97]. Thus, the HAWK manager must be trusted to not disclose private data of a smart contract, which is sent for the execution of the smart contract.

[4] Intel SGX is a set of central processing unit instruction codes that allows user-level code to allocate private regions of memory (referred to as enclaves). These enclaves are protected from processes running at higher privilege levels.

*C.2 User Unidentifiability vs. Throughput*

The less a network is controlled by a central authority and the more nodes participate in the network (i.e., given a high degree of decentralization), the vaguer is the identity of nodes. Therefore, *public-permissionless* distributed ledgers promise higher user unidentifiability than permissioned ones due to typically a higher number of nodes and a higher degree of decentralization. In contrast, a smaller, *permissioned* network with verified and identifiable nodes allows for higher throughput because faster consensus mechanisms can be used (e.g., PBFT). Nevertheless, unidentifiability can be improved by applying additional processes like mixing and the use of new keypairs for each transaction [102]. Yet, these processes create overhead due to preprocessing of each transaction, which results in extended transaction validation speed, and hence decreases throughput.

**D Opaqueness vs. Practicality**

*D.1 Node Controller Verification vs. Ease of Node Setup*

Verification of node controllers and their nodes' permissions (e.g., permissions to read data or to validate and commit new transactions) is required in permissioned DLT designs [69]. After permissions are granted to a node, the node can participate in (mostly, voting-based) consensus mechanisms (e.g., PBFT, PoA, or PoET; see Table 1). A public key infrastructure (PKI) with a trusted certification authority is often integrated to verify the nodes' identities and issue certificates to respective nodes [103]. However, a PKI produces additional efforts to obtain a certificate for the public private key pair and leads to the dependency on a trusted certification authority. Consequently, it becomes more complex to set up a node and participate in *permissioned* distributed ledger compared to *public-permissionless* DLT designs (e.g., Bitcoin or Ethereum) that only require to install an open source software client (e.g., *geth* or *parity* for Ethereum).

**E Performance vs. Performance**

*E.1 Block Creation Interval vs. Stale Block Rate*

Forkable and block-based DLT designs (e.g., Bitcoin or Ethereum) enable smaller block creation intervals. With smaller intervals, transactions can be faster appended to the distributed ledger; however, other nodes may not be aware of newly created blocks fast enough and may keep on working on already deprecated blocks. Ideally, each node would stop working on their blocks as soon as a new block is announced to save computational resources. However, nodes might receive newly created blocks too late due to propagation latency and keep working on an already stale block. Consequently, smaller block creation intervals increase stale block rate [104] and cause computational inefficiency.

**F Performance vs. Policy**

*F.1 Block Creation Interval vs. Degree of Decentralization*

In DLT designs where mining is performed to obtain a certain reward, a long block creation interval decreases the frequency of reward payouts and decreases the likelihood of rewards for individual miners. High variance in payments for miners makes it more likely that mining nodes will join mining pools[5] to increase the probability of receiving rewards [105,106]. However, the formation of mining pools decreases the degree of decentralization of a distributed ledger due to the collusion of miners in a mining pool [106].

*F.2 Confirmation Latency vs. Degree of Decentralization*

In *public-permissionless* DLT designs, participation of a high number of independent nodes (i.e., a high degree of decentralization) in consensus finding is required to protect the distributed ledger from malicious behavior and other Byzantine failures (see Figure 2). In turn, the number of nodes participating in the consensus mechanism negatively impacts the confirmation latency because agreement (e.g., *all nodes choose the same block*) and termination (e.g., *all nodes eventually choose a block*) in consensus finding cannot be reached at the same time in asynchronous systems [69,107]. As a solution, consistency and synchronicity must be relaxed in *public-permissionless* DLT designs to achieve liveness in a distributed ledger with a high degree of decentralization.

---

[5] A mining pool is a coalition of miners, who share mining rewards if one of these nodes receives the mining reward. Thus, the probability to receive some coins increases for each node.

*F.3 Throughput vs. Degree of Decentralization*

One major technological drawback inherent to current *public-permissionless* blockchains (e.g., Bitcoin or Ethereum) is a low throughput [108]. Such DLT designs are run by thousands of nodes. These nodes are operated by potentially malicious node controllers, which is why various DLT designs apply consensus mechanisms (e.g., Nakamoto Consensus) that reach consistency across multiple nodes with a comparably high (Byzantine) fault tolerance. However, most consensus mechanisms that allow for highly decentralized distributed ledgers only provide probabilistic finality to increase throughput by decreasing message complexity (see Table 1) [109]. In contrast, consensus mechanisms providing total finality (e.g., PBFT [41] or PoA [110]) can only include a comparatively small set of validating nodes due to their extensive communication overhead [107,109,111]. Hence, consensus mechanisms providing fast finality are commonly applied in *permissioned* DLT designs. Nevertheless, some consensus mechanisms for *public* DLT designs decrease the degree of decentralization to achieve an increase in throughput, for example, GoChain's PoR, which builds upon PoA and allows only selected organizations to run a validating node [36]. Meanwhile , there are consensus mechanisms that apply special derivates of PBFT to *public-permissionless* DLT designs (e.g., Tendermint [30] or EOS [31]), where a set of nodes is randomly chosen to reach consensus. However, this still centralizes decision making to a subset of nodes, decreasing the ledger's degree of decentralization. Thus, increased throughput comes at the cost of the degree of decentralization.

**G Performance vs. Practicality**

*G.1 Block Size vs. Transaction Fee*

Bitcoin has no mandatory transaction fees but allows for optional transaction fees (see trade-off *A.2*). However, miners could create huge blocks to receive transaction fees from as many transactions as possible, which eventually inhibits the distributed ledger from operating correctly. Generation of such huge blocks is prevented by introducing a maximum block size limits (e.g., in Bitcoin 1 MB per block or in Bitcoin Cash 2 MB per block). Nevertheless, such fixed block size limits flexibility in distributed ledgers because only limited data can be included in blocks. In contrast, Ethereum has no fixed maximum block size in favor of more flexibility (especially, when using smart contracts). However, Ethereum needed to solve the issue of potentially huge blocks, which is why transactions fees must be paid by any user of the Ethereum network.

**H Performance vs. Security**

*H.1 Confirmation Latency vs. Fault Tolerance*

Byzantine fault–tolerant consensus mechanisms come with an inherent trade-off between responsiveness and robustness [112]. Although enabling better responsiveness, allowing for forks in DLT harms robustness of a distributed ledger. To minimize the number of forks and to strengthen security, the targeted block creation interval must be set to a value that is large enough to minimize the stale block rate and small enough to confirm sufficient transactions within a certain period. If the block creation interval is set too short, the number of tolerable, malicious nodes decreases due to too many forks because new blocks are created before all nodes received the last valid block committed to the ledger; however, new blocks are faster confirmed for some nodes. On the other hand, if the block creation interval is set too long, confirmation latency increases because it takes more time to append a sufficient number of transactions so that it can be assumed that a transaction is committed to the ledger.

To cope with crashed nodes, weak synchronicity [112,113] is often applied, where the system designer makes timing assumptions on network delays to guarantee that the system will respond within a defined timeframe. A node is assumed as failed, if it did not respond within this timeframe (e.g., in PBFT and consensus mechanisms that adapt PBFT) [113]. In DLT designs such as Bitcoin and Ethereum [23], where the number of nodes is unknown, the timing assumption is expressed by the targeted average block creation interval [39], which prevents nodes from working too long on already stale blocks. Due to the assumption of weak synchronicity in the consensus mechanism, the targeted block creation interval strongly depends on the assumed block propagation time [114]. Timing assumptions (or block creation intervals) must be well balanced. If the timing assumption is too short, too many nodes would be considered as failed, which weakens robustness (e.g., fault tolerance) of the underlying security model. If the timing assumption is too long responsiveness decreases [112].

*H.2 Throughput vs. Consistency*

For the DLT concept blockchain, it was found that an increased block size can increase throughput because more transactions can be included in a block [11]. An increased size of data packets (i.e., blocks or transactions) comes with a longer propagation delay [11,69,115], which results in a longer state of inconsistency between nodes in a distributed ledger [72,116]. For Bitcoin and Ethereum, it was found that the percentage of created blocks that are successfully committed to the blockchain's main chain becomes low as the block size (and consequently the block propagation delay) increases [11]. Consequently, the stale block rate increases and nodes have inconsistent views on the ledger until the forks are resolved. Such inconsistent states, in turn, facilitate successful attacks (see Section 2.2). Thus, forkable DLT designs based on PoW can only improve throughput by degrading consistency (and increasing vulnerability) [117].

In BlockDAGs and TDAGs, throughput and scalability are usually much higher than in blockchains because the number of transactions per second is not bound to the block size, the block creation interval (due to relaxed consistency), and the requirement of being eventually added to the main chain. Instead, BlockDAGs and TDAGs require the blocks or transactions to be linked to previous blocks or transactions (e.g., IOTA, RChain, or soteriaDAG), so that nodes do not necessarily store an identical version of the ledger. Scalability in terms of increasing or decreasing throughput is theoretically infinite. However, such systems are much more complex than blockchains because they often aim to be fully asynchronous; and the process of converging toward a consistent state among all nodes is mostly not deterministic (e.g., in IOTA).

*H.3 Throughput vs. Fault Tolerance*

In blockchains, the requirement for high throughput is predominantly met by applying finality-preserving consensus mechanisms where a small set of nodes participates in the transaction and block validation process (e.g., PBFT [41]). In such consensus mechanisms, the number of nodes *n* determines the message complexity for the synchronization of the node states in $O(n^2)$. Due to the exponential increase in message complexity in contrast to probabilistic consensus mechanisms, finality comes at the cost of the degree of decentralization of the DLT design and its fault tolerance. In the case of *public-permissionless* DLT designs for the use of cryptocurrencies, fault tolerance is, for instance, predominantly prioritized above all other DLT properties such as performance and flexibility. For example, the Bitcoin blockchain achieves an average throughput of only 7 transactions per second and a transaction takes on average of 10 min to be committed [23]. However, Bitcoin is fault tolerant to up to 50 % of fraudulent validating nodes [23]. In contrast, PBFT in Hyperledger Fabric achieves a moderate throughput of three thousand transactions per second but tolerates only $f \leq \frac{|R|-1}{3}$ faulty nodes in a set of validating nodes in the distributed ledger (*R*) [41].

*H.4 Throughput vs. Integrity*

Increased block size can increase throughput because more transactions can be included in a block [11]. Bandwidth [100] and the current size of a block strongly influence the block propagation delay. Thus, the increased throughput comes with longer block propagation delays because more transactions are included in a block. However, longer block propagation delays increase the probability of forks [69], which threaten integrity and facilitate successful partition-based attacks on the distributed ledger [118] (e.g., selfish-mining [104,119], long-range attacks [54], bribery attacks [52]; see Section 2.2). To preserve integrity, the block creation interval must be adjusted in concert with the block size because a longer block creation interval mitigates the occurrence of forks and resulting attacks in blockchains [54]. Nevertheless, long block creation intervals also decrease the number of blocks issued, ultimately decreasing throughput.

Furthermore, highly varying loads on the distributed ledger caused by variations in transaction frequency result in block size variations and variations in the block propagation delay [120]. Variations in the block propagation delay increase the probability of successful selfish-mining attacks, thereby, threatening integrity [16,53].

**I Policy vs. Flexibility**

*I.1 Degree of Decentralization vs. Maintainability*

Distributed ledgers and applications on DLT require efficient maintenance to allow adaption to changing requirements and to increase security of the DLT design [121]. *Public-permissionless* DLT designs enable a high degree of decentralization, thereby, supporting unidentifiability because nodes do not need to be verified before joining the distributed ledger. Anybody is allowed to create new accounts. However, updates of software clients (e.g., *geth*) for a DLT design resulting from protocol changes must be accepted by the majority of nodes in the whole network to keep compatibility among the nodes after a hard fork and to prevent successful malicious behavior of nodes (see Section 2.1) [122]. The usually large number of nodes in *public-permissionless* distributed ledgers inhibits the introduction of mechanisms enforcing that nodes are kept up-to-date, which decreases maintainability of a DLT design [69]. Thus, *public-permissioned* DLT designs come with the costs of lower maintainability. In contrast, *private-permissioned* DLT designs are better maintainable because each node is verified and node controllers can be contacted directly.

**J Policy vs. Security**

*J.1 Degree of Decentralization vs. Integrity*

As the network size increases in *public-permissionless* DLT designs, it becomes unlikely that participating node controllers have the same (malicious) intentions or even know each other. Hence, the degree of decentralization increases (see Figure 2) and the presence of a group of nodes with shared interests that takes control of the distributed ledger becomes unlikely (e.g., by gaining a majority of, for example, 51 % of the overall hashing power). This strengthens the integrity of the respective DLT design.

In contrast, *private* DLT designs typically incorporate a small number of identifiable (trusted) nodes operated by verified node controllers, thus, decreasing the degree of decentralization and threatening integrity. Each node of such a *private* DLT design has an increased influence in the distributed ledger, which increases vulnerability, for example, toward the blockchain anomaly (cf. Section 2.2) [27]. Thin nodes, which only store parts of the distributed ledger, must assume that validating nodes verify all blocks and follow a working incentive mechanism when creating blocks. Otherwise thin nodes risk to accept invalid transactions [123]. End-users who only retrieve data from the distributed ledger and do not verify the distributed ledger's integrity on their own (e.g., using simple payment verification) cannot be sure that the distributed ledger's transaction history has not been tampered with [124].

In *permissioned* DLT designs, where only a subset of nodes is permitted to validate transactions and issue new blocks, the degree of decentralization of a distributed ledger decreases. However, *permissioned* DLT designs (and small *private-permissionless* DLT designs) can make use of consensus mechanisms that preserve total finality (e.g., PBFT). After total finality has been reached among the validating nodes, committed transactions cannot be retroactively changed. Hence, the trade-off between degree of decentralization and integrity predominantly refers to DLT designs that make use of probabilistic finality.

**K Policy vs. Practicality**

*K.1 Degree of Decentralization vs. Transaction Fee*

The degree of decentralization comes at the cost of higher transaction fees due to the applied consensus mechanism. In DLT designs that employ consensus mechanisms with probabilistic finality and rely on leader election based on PoW (e.g., Nakamoto Consensus), the degree of decentralization is important to ensure integrity of the stored data. To achieve a high degree of decentralization, *permissionless* DLT designs are characterized by extreme openness for new nodes. Arbitrary nodes can join the distributed ledger to participate in consensus finding and to validate transactions—without requiring permissions. As computations on blockchains are performed on each node, the total computational effort for the distributed ledger increases with an increasing number of nodes while the average transaction rate is constant. To compensate computational efforts, such DLT designs apply an economic incentive mechanism that rewards nodes for their share of resources [69]. The economic rewards result in a pricing structure that expects transaction issuers to pay transaction fees for the transaction processing and respective computational efforts.

**Table 6: Overview of identified trade-offs between DLT characteristics for the generated archetypes and exemplary DLT designs**

| Trade-Offs between DLT Characteristics | | | Archetypes | | | | | | Exemplary DLT Designs | | | | | | |
|---|---|---|---|---|---|---|---|---|---|---|---|---|---|---|---|
| A | B | ID | Flexibility | Opaqueness | Performance | Policy | Practicality | Security | Bitcoin | Ethereum | Hyperledger Fabric | RChain | soteriaDAG | IOTA | Nano |
| Block Size (+) | Transaction Fee (-) | G.1 | A | - | - | - | B | A | B | A | A | AB | B | - | - |
| Block Creation Interval (+) | Degree of Decentralization (+) | F.1 | B | B | A | - | A | B | B | B | A | AB | AB | - | - |
| | Stale Block Rate (-) | E.1 | - | - | B | - | A | B | B | A | A | A | AB | - | - |
| Confidentiality (+) | Integrity (+) | M.1 | B | A | B | B | B | A | B | B | A | AB | B | - | B |
| Consistency (+) | Availability (+) | M.2 | B | - | A | A | B | B | B | B | A | B | B | B | B |
| Confirmation Latency (-) | Fault Tolerance (+) | H.1 | A | - | A | B | A | B | B | B | A | A | A | A | AB |
| | Degree of Decentralization (+) | F.2 | B | B | A | A | A | B | B | B | A | A | A | A | AB |
| Degree of Decentralization (+) | Integrity (+) | J.1 | A | A | B | B | B | A | A | A | B | A | A | B | AB |
| | Maintainability (+) | I.1 | B | A | B | B | B | B | A | A | B | A | A | A | A |
| | Transaction Fee (-) | K.1 | A | A | B | - | B | A | A | A | B | AB | A | B | AB |
| Strength of Cryptography (+) | Support for Constrained Devices (+) | L.1 | B | A | - | A* | B | A | A | A | - | A | A | B | B |
| Maintainability (+) | Availability (+) | B.1 | A | - | A | A | B | B | B | B | A | B | B | B | B |
| | Integrity (+) | B.2 | A | - | A | A | B | B | B | B | A | B | B | X | B |
| Node Controller Verification (+) | Ease of Node Setup (+) | D.1 | B | B | A | A | B | A | B | B | A | B | B | B | B |
| Turing-complete Smart Contracts (+) | Confidentiality (+) | B.3 | A | B | - | - | A | B | - | A | AB | AB | X | X | X |
| | Resource Consumption (-) | A.1 | A | - | B | - | B | - | - | A | AB | AB | B | B | B |
| | Transaction Fee (-) | A.2 | A | - | B | - | B | - | B | A | AB | A | B | B | B |
| | Transaction Validation Speed (+) | A.3 | A | - | B | - | B | - | B | A | A | A | B | B | B |
| Throughput (+) | Consistency (+) | H.2 | A | - | B | B | A | B | B | B | A | AB | AB | A | AB |
| | Degree of Decentralization (+) | F.3 | A | B | A | - | A | B | B | B | A | AB | AB | AB | AB |
| | Fault Tolerance (+) | H.3 | A | - | A | B | A | B | B | B | A | AB | A | - | AB |
| | Integrity (+) | H.4 | A | - | A | B | A | B | B | B | A | AB | AB | A | AB |
| User Unidentifiability (+) | Resource Consumption (+) | C.1 | B | A | - | - | B | A | - | - | X | B | B | B | B |
| | Throughput (+) | C.2 | B | A | B | B | B | B | A | A | B | B | A | B | B |
| | | | | | | | | | Blockchain | | | BlockDAG | | TDAG | |

A: DLT characteristic A outweighs DLT characteristic B
AB: DLT characteristic A and B are both achieved (trade-off avoided by other means)
-: Trade-off not applicable
(+): DLT characteristics is aimed to be high

B: DLT characteristic B outweighs DLT characteristic A
X: Neither DLT characteristic A nor B are achieved (neither characteristic seems to be a design goal for the ledger)
*: only signatures
(-): DLT characteristic is aimed to be low

In contrast, various voting-based consensus mechanisms (e.g., PBFT or EOS's PoS) do not require transaction fees but allow only for a low degree of decentralization since they are unsuitable for a large number of validating nodes compared to, for example, Nakamoto consensus. The low degree of decentralization results from the fact that the applied consensus mechanisms require each node to agree on a certain state to reach total finality before a new transaction is committed to the distributed ledger. In addition, such finality-preserving consensus mechanisms are usually less costly than PoW-based consensus mechanisms in overall costs due to lower consumption of computational resources [107].

**L Security vs. Practicality**

*L.1 Strength of Cryptography vs. Support for Constrained Devices*

The strength of cryptography of a DLT design is dependent on the degree of security reached by algorithms for the generation of public-private key pairs (e.g., to secure authentication), for content encryption (e.g., to protect confidentiality), and for hash value calculation (hashing). For public key encryption it is important that the key pairs are unique and cannot be guessed. The algorithm's time complexity is important for encrypting/signing and decrypting/verifying data. In addition to time complexity in public key encryption, the applied hash algorithm implies a likelihood for collisions [125]. Low collision likelihood is desirable, which is why more secure hashing and key generation approaches are required (e.g., more bits for the output hash). However, an increased strength of cryptography requires more computational resources, such as random access memory and storage memory [126]. Thus, constrained devices such as microcontrollers can only hardly handle resource-intensive cryptography [126,127].

**M Security vs. Security**

*M.1 Confidentiality vs. Integrity*

To improve confidentiality, DLT designs are often implemented in a private network, where only select nodes can join (i.e., *private* DLT designs), for example, a private Ethereum blockchain or Hyperledger Fabric. However, a small number of known nodes makes it easier to have detailed information on the network topology. Access to a detailed network topology facilitates initiation of targeted delays in the communication between nodes because the data flow is known [49]. Thus, the probability for successful partition-based attacks [49] increases in private, forkable DLT designs such as a private Ethereum blockchain, which increases the likelihood for violations of a distributed ledger's immutability. Increased vulnerability for immutability violations reduces the integrity of a distributed ledger.

*M.2 Consistency vs. Availability*

Distributed systems theory reveals a trade-off between consistency and availability—the CAP Theorem [115,128]. This trade-off also persists in the field of DLT and is caused by latency in block propagation, for example, due to big block sizes or network failures. The larger the number of nodes that must receive new transactions, the longer the distributed ledger is in an inconsistent state. The larger the number of nodes of a distributed ledger, the more time it takes until each node has received the new block. However, many replications of the data stored on the distributed ledger increases availability. Thus, there is a trade-off between high availability and fast consistency.

## 4.2 DLT Design Archetypes for Applications on DLT

We introduce six archetypes of DLT designs to illustrate and to consolidate the previously presented trade-offs. Figure 3 illustrates the identified trade-offs on the DLT-property layer. The archetypes indicate benefits and drawbacks for applications on DLT that result from the choice and configuration of a DLT design, which is optimized toward a certain DLT property. Table 6 gives an overview of the identified trade-offs between DLT characteristics for the archetypes and exemplary DLT designs.

*Flexibility Archetype.* The flexibility archetype is designed to achieve high degrees of freedom in deploying applications on and customizing a DLT design. The flexibility archetype for a DLT design is predominantly characterized by the following five DLT characteristics: support for *Turing-complete smart contracts*, high *interoperability* with various external systems, a high *degree of maintainability*, a high *degree of decentralization*, and high *throughput*. Turing-complete smart contracts allow for the development of even complex applications on DLT. Furthermore, such expressive smart contracts are often required to enable interoperability between distributed ledgers, for example, by using simple payment verification [67]. In addition, the flexibility archetype must be efficiently maintained to allow for fast bug fixes or updates. In turn, efficient maintainability also requires an efficient change management, thus, governance mechanisms. Governance mechanisms pose a challenge in DLT designs with a high degree of decentralization (see trade-off *I.1*). However, distributed ledgers of the flexibility archetype should be capable of a high degree of decentralization to allow nodes to arbitrarily join and leave the distributed ledger; yet, the flexibility archetype should still achieve high throughput to be applicable in a variety of use cases, which may require high performance (e.g., sensor-based real-time monitoring of a production process).

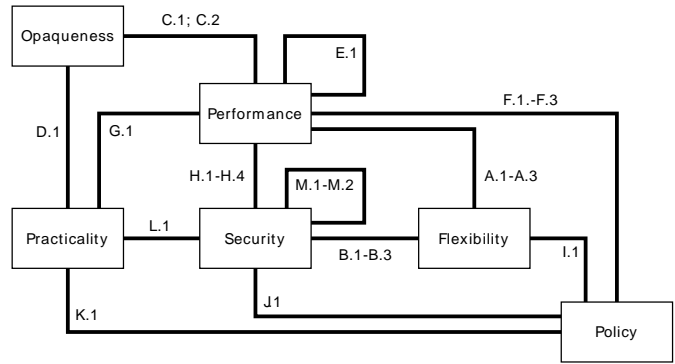

**Figure 3: Identified trade-offs between DLT properties**

The use of Turing-complete smart contracts in a distributed ledger with a high degree of decentralization comes at the cost of resource consumption (see trade-off *A.1*) and a slower transaction validation speed (see trade-off *A.3*). No fixed block size (or transaction size) should be introduced to allow for the deployment of smart contracts at any size. Accordingly, the flexibility archetype is likely to make use of transaction fees. The introduction of transaction fees also supports a high degree of decentralization because node controllers will receive rewards for their share of computing resources. The high degree of decentralization should be supported by a seamless ease of node setup. Ledgers of the flexibility archetype neglect integrity in favor of availability (see trade-off *B.1*) and maintainability (see trade-off *B.2*), which suggests that *public-permissioned* DLT designs could be employed to support efficient maintainability while ensuring a high availability. In the flexibility archetype, low resource consumption is preferred over user unidentifiability (see trade-off *C.1*), to allow for the use of devices that are constrained in their computational resources.

The focus on low resource consumption while supporting Turing-complete smart contracts suggests a need for thin nodes for constrained devices. Smart contracts should only be executed by full nodes in the distributed ledger but not by thin nodes. Furthermore, ledgers of the flexibility archetype are likely to apply sharding to achieve parallel execution of smart contracts and to achieve high throughput and a high degree of decentralization. A DLT design that aligns with the flexibility archetype has already been proposed for the *Serenity* update in the Ethereum protocol, which will divide the network into three distinct shards: transaction processing shard (*Main Chain*), consensus shard (*Beacon Chain*), and smart contract execution shard (*Sharding Chain*) [129]. Furthermore, DLT designs that strongly relax their consistency assumptions while still supporting Turing-complete smart contracts (e.g., RChain) align well with the flexibility archetype [25]. Applications deployed on ledgers of the flexibility archetype may become expensive to use if most of the logic is performed by smart contracts. In the development of applications on ledgers of the flexibility archetype, a balance must be maintained between the use of smart contracts and traditional programs to express and enforce program logic. A balance between use of smart contracts and traditional programs is also important for maintainability of the respective application. For example, functionalities that are updated frequently should not be stored on the distributed ledger or in a smart contract. because when the smart contract needs to be updated it is reset and starts from zero. If a smart contract is used, data stored in it (for example, a list of user accounts) should be kept in another smart contract that is only used for data storage and provision. By doing so, smart contract functionality can be maintained, while still keeping its current state.

*Opaqueness Archetype.* The opaqueness archetype is specialized to prevent use and operation of a DLT design to be tracked. This archetype is concerned with the achievement of a high degree of *user unidentifiability*, high *confidentiality*, a high *degree of decentralization*, and the absence of *node controller verification*. Confidentiality and user unidentifiability are the main requirements to be fulfilled by the opaqueness archetype. A high degree of decentralization is desired to support user unidentifiability. Node controller verification contradicts with the opaqueness archetype. Therefore, more seamless ease of node setup is preferred (see trade-off *D.1*).

Although ledgers of the opaqueness archetype are geared towards *ease of node setup* instead of *node controller verification*, a high *ease of node setup* and high ease of use, in general, may not be achieved in ledgers of the opaqueness archetype because the use of additional anonymization mechanisms is often recommended (e.g., use of the TOR network in Zcash). Such additional anonymization mechanisms may not be easy to comprehend and to apply for users and can pose a risk for user anonymity. To achieve user unidentifiability, additional processing of transactions is necessary (e.g., mixing or zero-knowledge proofs). These processes are time-consuming and require additional computational power, which slows down performance and can hardly be performed on constrained devices (e.g., microcontrollers or sensors). A high degree of decentralization increases unidentifiability but increases confirmation latency, thus, impedes consistency. From a policy point of view, auditability is impaired because transactions are not traceable and issuers and recipients of a transaction are not identified. The opaqueness archetype is likely to not offer Turing-complete smart contracts because smart contracts pose a threat to confidentiality and make it easier to identify transaction senders and receivers and to monitor their interactions (see trade-off *B.3*) [130]. Applications with a strong requirement for opaqueness should handle most of their advanced business logic off-chain because the opaqueness archetype will probably provide poor performance and flexibility. Due to the immutability of stored data, there is a threat of revealing encrypted content as technology evolves.

The opaqueness archetype aligns well with *public-permissionless* ledgers where multiple cryptographic techniques are applied (e.g., zero knowledge proofs) to make it as hard as possible to assign transactions to their senders and receivers or to reveal transaction contents. Popular representatives for the transparency archetype are Dash [37], Monero [131], and Zcash [59]. In Dash, additional fees must be paid if a transaction should be issued privately, which decreases practicality in terms of transaction fees. Dash still allows to view the transaction recipient. In Monero, ring signatures are applied to obfuscate the identity of involved parties [132]. However, Monero has been criticized for vulnerabilities that eventually make transactions traceable [133]. Although Zcash does not obfuscate IP addresses of clients, it is currently considered the most confidentiality-preserving DLT design (especially, when using it over the TOR network).

***Performance Archetype.*** The *performance archetype* is focused on allowing accomplishment of a given task on a distributed ledger under most efficient use of computing resources and time. Thus, the performance archetype is characterized by high *throughput*, low *confirmation latency*, low *resource consumption*, and a high *maintainability*. High throughput and confirmation latency can be achieved by keeping the number of validating nodes small (e.g., by using a *private* DLT design). A small number of validating nodes supports maintainability (see trade-off *I.1*) and can accelerate consistency among all nodes [115,128]. When deciding for a high degree of decentralization, consistency assumptions would need to be relaxed to achieve high throughput and scalability (e.g., in IOTA or RChain) [115,128].

DLT designs that align well with the performance archetype are not likely to support user unidentifiability or Turing-complete smart contracts to decrease the transaction processing time. Due to the short-targeted confirmation latency, the performance archetype will have lower fault tolerance. To accomplish fast confirmation latency with total finality, the performance archetype can be realized as a *private*(*-permissioned*) DLT design. Such *private* DLT designs come at the cost of availability [115,128], user unidentifiability (see trade-off *C.2*), the degree of censorship resistance, fault tolerance (see trade-off *H.3*), and integrity (see trade-off *J.1*). In private instantiations of the performance archetype, node controller verification is required, which decreases the ease of node setup (see trade-off *D.1*). If ledgers of the performance archetype should also scale to a huge number of nodes, a *public-permissionless* DLT design can be designed based on the blockDAG or the TDAG DLT concept, where consistency assumptions are relaxed compared to blockchains, which require a certain average block creation interval for synchronization. Such DLT concepts predominantly come with probabilistic consensus mechanisms that often have higher fault tolerance, but less integrity.

Multiple DLT designs targeting high performance that may follow a *private(-permissioned)* approach in blockchains (e.g., Hyperledger Fabric) or rather rely on blockDAGs (e.g., RChain) or TDAGs (e.g., IOTA), have been developed. To increase performance (especially, scalability of blockchains) sharding is applied, that is, multiple distributed ledgers exist in parallel and are connected with each other (e.g., in Zilliqa [134] or Wanchain [135]) [136,137]. Sharding requires interoperability between the DLT designs, which brings more complexity to the distributed ledger but also better maintainability of the particular distributed ledger. To sum up, applications requiring high-performance DLT designs have a limited degree of decentralization or increased complexity due to sharding. New consensus mechanisms are under development (e.g., $\varepsilon$-differential agreement), which scale proportional to the number of nodes in the network (e.g., seele [138]).

*Policy Archetype.* The *policy archetype* aims to offer a variety of abilities to guide and verify the correct operation of a DLT design. Thus, ledgers of the policy archetype are likely to make use of *node controller verification* to better *govern*, *maintain,* and *audit* the appropriate setup of their nodes. For efficient governance, various mechanisms are provided to users of ledgers of the policy archetype (e.g., standard smart contracts for voting). High maintainability in ledgers of the policy archetype allows for the introduction of updates, which makes the ledgers more flexible and capable to apply changes to the protocol to achieve compliance with targeted regulations or standards. To check compliance, auditability is important in the policy archetype. To audit data in the distributed ledgers, fast *consistency*, high *integrity*, and *non-repudiation* are of particular importance in the policy archetype, just as well as, *transaction content visibility* and *traceability*. Fast consistency among nodes contributes to less contradictions between the statements represented in the data stored on the nodes, which facilitates the auditing process. In addition, integrity, in particular, tamper-resistance, of once stored data helps to trace the history of logs (e.g., transfer of assets between users), which increases the reliability of audits. Finally, *non-repudiation* is important to be able to reliably map such logs to users in audits or governance.

The specialization of ledgers regarding the policy archetype predominantly comes at the cost of opaqueness-related DLT characteristics (i.e., *traceability*, *transaction content visibility,* or *user unidentifiability*) and, additionally, *confidentiality* (due to transaction content visibility) and *throughput* (see trade-off *H.2*). New regulations and standards are often introduced, and distributed ledgers must adapt to them to achieve compliance. Due to the targeted high level of integrity the ex post adaptation of a distributed ledger to reach compliance becomes challenging. For example, it is not possible to become compliant with the requirements imposed by the EU General Data Protection (GDPR) [139] if personal data is stored on a distributed ledger because GDPR demands for a possibility to completely delete personal user data. To increase flexibility in order to adapt applications on DLT to future regulations or standards, developers must carefully determine which data should be stored on chain or off chain [139,140]. For now, it remains unclear how to provide flexibility to become compliant with future regulations or standards and achieve a high level of integrity at the same time [e.g., 139]. Therefore, sensitive data should be predominantly stored off chain. Nevertheless, off-chain data stores are administrated by at least one trusted third party, which lowers the degree of decentralization of applications on DLT. In addition, external data needs to be kept confidential and available for the distributed ledger. Thus, reliable interoperability of DLT designs with oracles becomes important for the policy archetype [61]. Furthermore, the oracles themselves must also be compliant with the same laws and regulations.

Due to its strong requirement for transaction content visibility and traceability, the policy archetype is likely to be found as a *private* DLT design. All users are identifiable, and no unknown user is allowed to view the data stored on the distributed ledger, which increases confidentiality of the data. Furthermore, *private* DLT designs allow for better maintainability (see trade-off *I.1*) and faster consistency (see trade-off *M.2*) compared to *public-permissionless* DLT designs.

*Practicality Archetype.* Ledgers of the *practicality archetype* are designed to allow their users to achieve their goals with respect to social expectations on technology in everyday practice. Thus, DLT designs that align well with the practicality archetype offer high *throughput* and a low *confirmation latency* to achieve low response time (under consideration of transaction finalization), high *support for constrained devices*, and low *transaction fees*. In addition, the practicality archetype also provides Turing-complete smart contracts to allow for interoperability with other distributed ledgers [67] or non-DLT systems [61].

Despite the benefits of the practicality archetype, there are several drawbacks. Various *public-permissionless* DLT designs incentivize nodes to share resources with monetary mechanisms to reach a high degree of decentralization (see trade-off *K.1*). A high *degree of decentralization* and openness for new users of the distributed ledger is also targeted in ledgers of the practicality archetype. To make applications on DLT easily-usable by a large number of users that interact with the distributed ledger via a broad variety of devices including constrained devices such as sensors, a full replication of the ledger on each device should be avoided and ledgers of the practicality archetype should allow for the use of thin nodes. The requirements for high *scalability*, high *throughput,* and *fast confirmation latency* indicate that ledgers of the practicality archetype likely have poor *fault tolerance* (see trade-off *H.3*) and the *consistency* assumptions need to be relaxed (see trade-off *H.2*). Furthermore, the pragmatism comes at the cost of *confidentiality* (see trade-off *M.1*) and *user unidentifiability* (see trade-offs *C.1* and *C.2*).

Ledgers of the practicality archetype ensure that users do not need to have sound knowledge of DLT before using it, while allowing them to easily interact with the distributed ledger. Therefore, users of ledgers of the practicality archetype will usually not host their own node but will be offered other gateways to interact with the ledger through. The ledgers are operated by a consortium as (*private-* or *public-*) *permissioned* distributed ledgers. Because a private key cannot be recovered, users can no longer access their assets if they lose their private key. Therefore, the management of a public-private key pair should be made easy and secure for users, which is why the provision of secure tools for the organization of users' public and private keys is crucial. Exemplary DLT design that align with the practicality archetype are Hyperledger Fabric and EOS (see Table 1).

*Security Archetype.* The *security archetype* provides a high likelihood that the functioning of the distributed ledger and stored data will not be compromised. To achieve this goal, DLT designs are optimized toward high *availability*, high *fault tolerance*, high *integrity,* and high *confidentiality*, which may even include *user unidentifiability* to inhibit the mapping of data to identities. To achieve strong integrity and fault tolerance, the *degree of decentralization* should be high,

High availability can be achieved by adding numerous, physically distributed nodes to the distributed ledger, each maintaining a replication of the ledger. While large network size is comparably easy to achieve, achieving a high degree of decentralization is more challenging. However, the degree of decentralization is a focal requirement in the security archetype (e.g., trade-offs *B.1* or *J.1*). The degree of decentralization does not merely result from the DLT protocol. Instead, it predominantly depends on socio-technical phenomena, such as (ad-hoc) consortia of validating node controllers (e.g., mining pools; see Figure 2). Avoiding such consortia poses a particular challenge in the instantiation of the security archetype. Due to its high requirements for confidentiality, the security archetype is likely to require additional techniques that make the identification of users difficult (see opaqueness archetype). Such mechanisms come at the cost of increased resource consumption and less support for constrained devices (see trade-off *C.1*). The use of anonymization techniques (e.g., zero knowledge proofs) may cause serious security issues because it is hard to audit the distributed ledger due to decreased traceability (e.g., by applying mixing) and decreased transaction content visibility (e.g., by encryption).

Due to inherent trade-offs within the security archetype (e.g., trade-offs *M.1* and *M.2*), DLT designs that correspond to the security archetype are likely to either achieve *security through decentralization* or *security through permission.* The first aims to achieve a huge number of independent node controllers, which increases absolute fault tolerance (tolerable number of malicious nodes). The most prominent DLT design that follows the approach of *security through decentralization* is the Bitcoin blockchain [23]. The Bitcoin blockchain is highly available due to its high number of nodes and hardly offers potential for flawed smart contracts. Nevertheless, Bitcoin does not fulfill all the security-related DLT characteristics (e.g., confidentiality). To increase confidentiality, Zcash applied z̲ero-k̲nowledge s̲uccinct n̲on-interactive arguments of knowledge (zk-SNARKs) to obfuscate transaction senders (and receivers) and to impede transaction traceability. However, the added complexity of zk-SNARK has already caused a counterfeiting vulnerability for Zcash coins [141], which indicates that such techniques also allow for new vulnerabilities. *Security through permission* aims at limiting access to the distributed ledger to known users, which increases maintainability of the distributed ledger at the cost of its degree of decentralization (see trade-off *I.1*). Considering the advances in computer science, which could even break encryption (e.g. quantum computers), *security through permission* should achieve better confidentiality in the future.

# 5 Discussion

## 5.1 Principle Findings

Our research reveals twenty-four trade-offs (see Table 6) based on forty identified DLT characteristics (see Table 5), which we grouped into six DLT properties (see Table 4). The diversity of the identified DLT characteristics from purely technical (e.g., *strength of cryptography* in *security*) to social (e.g., *degree of decentralization* in *policy*) highlights the complexity of DLT. Among the abstracted trade-offs, the DLT properties *performance* and s*ecurity* each exhibit the most trade-offs (11) (see Figure 3 and Table 6). Purely performance- or security-oriented DLT designs appear to be challenging and maybe even impossible to be developed because of trade-offs between DLT characteristics within the respective DLT property (see *performance* and *security archetype*).

The consolidation of the identified trade-offs between DLT characteristics into archetypes elucidates that it is not possible to develop a one-size-fits-all DLT design that fulfills all requirements of each application. Thus, application designers will have to wisely choose a DLT design when aiming to develop viable applications on DLT. Nevertheless, the derived archetypes partially support each other (e.g., opaqueness and security), while others contradict (e.g., performance vs. security). This phenomenon can also be seen as an indicator of compatibility between DLT designs. The preference of one DLT characteristic over another in a trade-off is critical. For example, the security archetype can benefit from certain features of DLT designs leveraging the opaqueness archetype, these DLT designs can even benefit from each other if they were combined. To jointly use DLT designs that have made contradicting decisions in the trade-offs, interoperability between DLT designs is required because otherwise they could not synchronize. For example, DLT designs that are matched to the performance archetype are difficult to link to DLT designs that are matched to the security archetype.

Several identified trade-offs are inherent to distributed systems, for example, those related to the CAP theorem [10,115] (see trade-off *M.2*) or the FLP impossibility [142] (see trade-off *H.1*). However, the implications of these trade-offs on DLT designs differ from commonly used distributed databases, where mostly a known number of nodes is employed, and consensus mechanisms are predominantly crash fault-tolerant but not Byzantine fault-tolerant (e.g., Paxos [143] or Raft [144]). In DLT, implications of such design decisions do not only impede consistency of the distributed ledger but may have serious (financial) consequences that result from inconsistencies (e.g., form forks) and successful attacks that make use of inconsistencies to weaken a distributed ledger's integrity (e.g., double spending).

The analyzed literature showed a trend toward preferring *private* DLT designs over *public* DLT designs for industrial applications. The shift from pure decentralization through a high degree of openness (e.g., Bitcoin and Ethereum) toward more centralization seems largely motivated by improvements in performance due to the employment of faster consensus mechanisms, and enhancement of confidentiality due to restricted access to the ledger. Nevertheless, there is much criticism on this shift, specifically, because it contradicts with DLT's original philosophy. Similarly, this shift aligns with our observation that most trade-offs between DLT characteristics are related to the DLT characteristics *security* or *performance*. *Private* DLT designs come with various advantages compared to *public* DLT designs with respect to practicality but tend to reach a low degree of decentralization compared to *public* DLT designs. Furthermore, *private* DLT designs strongly require universal *interoperability* with other DLT designs or external services to prevent being caught on a *'blockchain island'* [145–148]. Furthermore, there are various DLT designs of the DLT concepts BlockDAG and TDAG that incorporate rather loose structures based on DAGs (e.g., IOTA, Nano, or RChain). These DLT designs relax consistency assumptions in favor of high throughput and consensus algorithms that consume less resources compared to most blockchains (e.g., Bitcoin or Ethereum). Nevertheless, support of Turing-complete smart contracts is not yet widespread, which limits flexibility of several DAGs.

DLT designs such as Nano and RChain even advance the use of DAGs in DLT by not rigorously applying the concept of replicated state machines in favor of less storage consumption. In Nano, only personal transactions are, for example, stored on the terminal device. In RChain, DLT is, for example, strongly connected with peer-to-peer file sharing, which is also targeted in other projects such as Ethereum's Swarm. The integration of such external data allows for the extensive use of (unreliable) data sources and may announce the birth of a new generation of peer-to-peer systems in general, where DLT might take an important position by allowing for asset exchanges and tamper-resistant proofs of such asset exchanges.

## 5.2 Main Lessons to be Learned

Our work provides diverse contributions to research and practice. Regarding the latter, practitioners obtain deep insights into viability of DLT designs for applications on DLT and their possible impacts on organizations. Our work supports the decision making for a DLT design and its later configuration to use for applications on DLT under consideration of application requirements and DLT characteristics. The overview of DLT characteristics and DLT properties supports practitioners in defining requirements for DLT designs that must be considered in the requirements engineering process to ensure viability of applications on DLT. The derived trade-offs between DLT characteristics and the generated archetypes suggest potential benefits and drawbacks for applications on DLT, which can be assessed before starting to develop the application (see Table 6). Such assessments eventually facilitate avoidance of unsuitable DLT designs and consequent waste of resources. To understand causes of such drawbacks for applications on DLT, the described trade-offs between DLT characteristics provide rationale (see Section 4.1). Careful DLT design selection and application development becomes crucial to ensure that DLT's unique advantages can be achieved, ultimately, pushing DLT from a hype to a critical information infrastructure [149] for future businesses and societies.

Our synthesis of the four previously disconnected research streams on DLT (*description, analysis, application,* and *guidance*) bridges different research streams in DLT, thus, contributes to a more holistic view of DLT. The *description* research stream on DLT is consolidated in this work with a strong focus on applications on DLT. Our classification of DLT characteristics can be used to generate a common understanding of important terms in the field of DLT and their technical dependencies across research fields, such as economics or computer science. Our findings support the development of comprehensive models and simulations of DLT designs, which has only partially been approached so far [e.g., 150,151]. The results of such *analyses* (e.g., formalization of dependencies between DLT characteristics) will elucidate assessments of the influence of the identified trade-offs between DLT characteristics on applications on DLT. Research on the *application* of DLT is supported in decision making for a particular DLT design. Research on business process innovation using DLT, for example, can draw from the trade-offs and archetypes to discuss possible negative effects of the integration of DLT. Furthermore, we contribute to research on software engineering and requirements engineering in distributed systems since a holistic view on non-functional requirements can be obtained. Finally, we support the DLT research stream *guidance* by introducing the archetypes of DLT design. The archetypes of DLT design form a fundament for a preselection of DLT designs for applications and can support the selection of an appropriate DLT design [e.g., 71] to make the selection of a DLT design more efficient.

## 5.3 Limitations

Nevertheless, our study comes with limitations. DLT characteristics and DLT properties were identified in a literature review in the field of DLT. Analyzed DLT concepts are limited to already published scientific articles and mainly focused on blockchain. Therefore, we limit our overview of DLT characteristics to those of particular interest in extant research on DLT for the development of applications. The DLT characteristics and related trade-offs are also corroborated by multiple whitepapers of DLT designs such as Bitcoin [31], Ethereum [36], or soteriaDAG [26] (see Table 6). Most of the analyzed research articles in the application research stream developed applications on Bitcoin, Ethereum, or Hyperledger Fabric. This makes our work, the trade-offs in particular, only partially generalizable to other DLT designs. We tried to overcome this limitation by including feedback from several DLT experts in a survey and illustrate the generalizability of the identified trade-offs between DLT characteristics by applying them to DLT designs of all three DLT concepts known so far. There are preliminary approaches for the analysis and formalization of DLT concepts for the development of frameworks for the simulation of DAGs [e.g., 151]. However, we could not identify trade-offs between DLT characteristics that are specific for BlockDAGs and TDAGs based on the reviewed literature. While we analyzed dependencies between DLT characteristics, we predominantly focused on potential negative effects and resulting trade-offs. We acknowledge that dependencies might also lead to synergistic and positive effects.

## 5.4 Future Research

We identified multiple very influential conditions that impact the strength of certain dependencies or even the presence of trade-offs between DLT characteristics such as the applied consensus mechanism or the use of additional services such as mixing (e.g., regarding the trade-off *unidentifiability vs. throughput* (C.1)). Thus, researchers of the *analysis* research stream should conduct measurements to quantify the identified trade-offs between DLT characteristics under different conditions. The analysis should include other DLT concepts than blockchain to reveal dependencies between DLT characteristics for different DLT concepts. Such analyses support the quantification of the dependencies between DLT characteristics, which can be used to quantify the influence of the particular trade-offs between DLT characteristics. Quantified trade-offs would support the development of decision support systems for the selection of DLT designs for applications in the *guidance* research stream. Based on a quantified model of the trade-offs, monitoring-systems for distributed ledgers can be developed, which can use the generated trade-offs to predict the behavior of a distributed ledger. Researchers of *applications* on DLT can further investigate how to design decision-support and monitoring applications for DLT.

As the identified archetypes inhibit simultaneous optimization of certain DLT characteristics due to DLT-inherent trade-offs, interoperability between DLT designs (cross-chain technology) turns out an important avenue for future research in the field of DLT to overcome prevalent issues in DLT [e.g., scalability, throughput, or lack of smart contracts 67,152]. Research on cross-chain technology is still in its infancy and is, for example, concerned with the transfer of assets from one distributed ledger to another [145]. Cross-chain technology can increase flexibility of DLT designs and might help to mitigate the inherent trade-offs through multi-chain networks, which are allow for the benefits of any DLT design while avoiding the drawbacks through clever cross-chain technology.

## 5.5 Conclusion

Reading this manuscript underpins the notion that there cannot be a one-size-fits-all DLT design due to dependencies and consequent trade-offs between DLT characteristics. Since it is difficult to consider all the trade-offs and their particular impact at once, this manuscript introduces archetypes of DLT designs and illuminates twenty-four prevalent trade-offs in DLT designs. The archetypes of DLT designs support practitioners in understanding causes of benefits and drawbacks of particular applications on DLT, which will result from the selected DLT design. The trade-offs and their consolidation into archetypes make the challenges inherent in the configuration of a DLT design more transparent for developers and are useful to prevent wrong decisions before choosing a DLT design. Beyond Blockchain, our survey article suggests that the true potential of DLT might lie in decentralization of applications that are not as restrictive as Bitcoin transactions while still empowering the individual. We wrote this survey article in the hope that it will be helpful to successfully navigate this future transformation.

## ACKNOWLEDGEMENTS
This work was carried out in the scope of the project COOLedger (Helmholtz Association of German Research Centers: HRSF-0081, Russian Science Foundation: Project No. 19-41-06301). We thank all participants of the empirical studies that contributed to this work. In particular, we would like to thank Mikael Beyene and Konstantin Pandl for taking the time to discuss the trade-offs based on the archetypes and selected DLT designs.